\documentclass[showpacs,prl,twocolumn,aps,superscriptaddress,preprintnumbers,letterpaper]{revtex4}

\usepackage{amsmath,amssymb}
\usepackage{epsfig}
\usepackage{graphicx}
\usepackage{amsmath}
\usepackage{amsfonts}
\usepackage{epstopdf}
\usepackage{color}

\newcommand{\be}{\begin{equation}}
\newcommand{\ee}{\end{equation}}
\newcommand{\bear}{\begin{eqnarray}}
\newcommand{\eear}{\end{eqnarray}}

\newcommand{\ba}{\begin{array}}
\newcommand{\ea}{\end{array}}

\begin{document}

\title{Deconfinement as an entropic self-destruction:\\ 
a solution for the quarkonium suppression puzzle?}

\author{Dmitri E. Kharzeev}
\affiliation{Department of Physics and Astronomy, Stony Brook University, Stony Brook, New York 11794-3800, USA}
\affiliation{Department of Physics, Brookhaven National Laboratory, Upton, New York 11973-5000, USA}

\date{\today}

\begin{abstract}
The entropic approach to dissociation of bound states immersed in strongly coupled systems is developed. In such 
systems, the excitations of the bound state are often delocalized and characterized by a large entropy, so that the bound state is strongly entangled with the rest of the statistical system.   If this entropy $S$ increases with the separation $r$ between the constituents of the bound state, $S = S(r)$, then the resulting {\it entropic force} $F = T\ {\partial S}/{\partial r}$ ($T$ is temperature) can drive the dissociation process. As a specific example, we consider 
the case of heavy quarkonium in strongly coupled quark-gluon plasma, where lattice QCD indicates a large amount of entropy associated with the heavy quark pair at temperatures $0.9\ T_c \leq T \leq 1.5\ T_c$ ($T_c$ is the deconfinement temperature); this entropy $S(r)$ grows with the inter-quark distance $r$. We argue that the entropic mechanism results in an anomalously strong quarkonium suppression in the temperature range near $T_c$. This {\it entropic self-destruction} may thus explain why the experimentally measured quarkonium nuclear modification factor at RHIC (lower energy density) is smaller than at LHC (higher energy density), possibly resolving the ``quarkonium suppression puzzle" -- all of the previously known mechanisms of quarkonium dissociation operate more effectively at higher energy densities, and this contradicts the data. Moreover, we find that near $T_c$ the entropic force leads to delocalization of the bound hadron states; we argue that this delocalization may be the mechanism underlying deconfinement.
\end{abstract}


\maketitle

\setcounter{footnote}{0}



\section{Introduction} 
Entropy is one of the key concepts in modern science, with applications that 
far transcend the boundaries of its native thermodynamics.
For example, the entropy of the black hole \cite{Bekenstein:1972tm,Hawking:1974sw} was instrumental in understanding the interplay of quantum mechanics and gravity, and subsequently led to the holographic \cite{'tHooft:1993gx,Susskind:1994vu} gauge/gravity correspondence \cite{Maldacena:1997re,Gubser:1998bc,Witten:1998qj}. The entanglement entropy provides a non-local order parameter of topological order in strongly coupled systems \cite{Nishioka:2009un}. 
The absence of entropy production serves as a stringent constraint on anomaly-induced non-dissipative transport \cite{Kharzeev:2011ds}, fixing most of the transport coefficients in chiral magnetohydrodynamics \cite{Son:2009tf,Kharzeev:2013ffa}. 
Quite often, the considerations based on the entropy allow to understand the behavior of complex systems with complicated dynamics not amenable to microscopic treatment.  
\vskip0.2cm
In particular, if the entropy $S$ of a composite system depends on the coordinate $r$ of a constituent, it is useful to introduce the notion of the {\it entropic force} with magnitude  
\be\label{entf0}
F(r) = T \frac{\partial S}{\partial r}  .
\ee
The entropic force does not describe any additional fundamental interaction; instead, it is an {\it emergent} force that stems from multiple interactions driving the system, in accord with the second law of thermodynamics, towards the state with a larger entropy. The entropic force was originally introduced \cite{Meyer} to explain the elasticity of 
polymer strands in rubber. The rubber polymer strands are long, and when stretched, possess smaller entropy than in the ground state where their motions are unrestricted. The stretched polymers thus tend to contract to the ground state, and this causes a macroscopic entropic force resulting in the contraction of the stretched rubber band. The underlying fundamental interactions are of course electromagnetic, but the notion of entropic force allows to bypass the consideration of complicated microscopic dynamics.  
\vskip0.2cm
The notion of entropic force offers a simple alternative way of deriving kinetic theory; for reader's convenience, we will outline below an entropic approach to diffusion developed by Neumann \cite{Neumann}. It has been proposed by Verlinde \cite{Verlinde:2010hp} that the entropic force may play a much more profound role in physics, being responsible for gravity. This intriguing idea is a subject of a lively controversy, and will not be discussed here. We will restrict ourselves to statistical physics where the notion of the entropic force has been firmly established. 
Moreover, the entropic force is even put to practical use in {\it entropic self-assembly}, where nano-particles arrange themselves in a desired pattern that maximizes their entropy; see e.g. \cite{self-assembly}.
\vskip0.2cm
In this paper, we address the behavior of bound states in QCD matter at finite temperature. We will argue that the process of deconfinement can be viewed as an {\it entropic self-destruction}, when  bound hadron states are driven towards a delocalized state that maximizes the entropy of the system. This delocalization occurs at temperatures around the deconfinement temperature, and may be considered as an entropic representation of the deconfinement itself. Specifically, we consider the dissociation of heavy quarkonia in quark-gluon plasma (originally proposed as a signature of deconfinement in \cite{Matsui:1986dk}) within this entropic framework. In this case, the increase of the entropy associated with the heavy quark-antiquark pair with the inter-quark distance has been observed in lattice QCD \cite{Kaczmarek:2002mc,Petreczky:2004pz,Kaczmarek:2005zp}, and so the entropic force (\ref{entf0}) should be present. 
\vskip0.2cm
The physical reason for this increase of the entropy with the inter-quark distance is likely the abundance of the physical states that become available for the separating heavy quarks -- while at short distances the color dipole moment of the pair is small and it decouples from the medium, at larger distances the heavy quarks may form extended bound states characterized by a larger entropy. This picture is supported by the recent lattice results \cite{Bazavov:2014yba} indicating that close to the crossover transition the charmed degrees of freedom can no longer be described using an uncorrelated gas of known hadrons. In string picture, this increasing entropy is associated with the entropy of a ``long string" \cite{Kogut:1974ag, Aharonov:1987ah, Deo:1989bv,Kalaydzhyan:2014tfa,Hashimoto:2014xta} connecting the heavy quark pair; the condensation of long strings (equivalent to a black hole formation \cite{Hanada:2014noa}) describes a deconfined phase.  The condensation of long strings (or "string nets") can also describe the topological phases in condensed matter systems \cite{Levin:2004mi}, implying an interesting cross-disciplinary connection. Indeed, it has been proposed recently that QCD matter can be viewed as a topological phase \cite{Zhitnitsky:2013hs}.

\vskip0.2cm
By (\ref{entf0}), the increase of the entropy with the quark-antiquark distance leads to the entropic force that points outward and can induce the self-destruction of the bound state. Below we will find that the resulting delocalization of heavy quarks, and thus the quarkonium suppression rate, is maximal near the deconfinement transition temperature. This provides a possible explanation for the puzzling energy dependence of the heavy quarkonium nuclear modification factor observed at RHIC \cite{Adare:2006kf} and LHC \cite{Abelev:2013ila}: even though the density of produced matter is higher at LHC than at RHIC, the nuclear modification factor at LHC appears larger than at RHIC. 
It has been pointed out \cite{Satz:2013ama} that an appropriate measure of charmonium suppression is the ratio of the hidden-to-open charm, and not the nuclear modification factor (which is the normalized ratio of nucleus-nucleus and $pp$ charmonium production cross sections). Even so, to reconcile the increase of the charmonium nuclear modification factor at the LHC with the stronger suppression expected in conventional scenarios would require a large increase in the production of open charm at the LHC, which would be a puzzle in itself; the forthcoming data on open charm production at small transverse momentum will clarify the situation. 
A possible explanation of the charmonium suppression puzzle is the heavy quark recombination \cite{BraunMunzinger:2000px,Thews:2000rj}, see \cite{Andronic:2014zha} for a recent review. However, here we propose an alternative explanation linked to the nature of deconfinement transition.  
\vskip0.2cm

\section{Entropic view on Einstein's diffusion} 

Let us begin by summarizing the entropic approach to diffusion proposed by Neumann \cite{Neumann}; see \cite{Roos} for a recent discussion. 
Consider a particle released at the origin $r=0$. The number of states for the particle at distances between $r$ 
and $r + dr$ is proportional to the volume $dV(r) = 4 \pi r^2 dr \equiv \Omega(r) dr$, and the corresponding $r$-dependent part of the entropy  is 
\be\label{entdif}
S(r) = k \ln \Omega(r) = 2 k  \ln r + const ;
\ee
where we wrote explicitly the Boltzmann constant $k$.
The resulting entropic force is 
\be\label{entf}
F(r) = T \frac{\partial S}{\partial r} = \frac{2 k T}{r} .
\ee
In a viscous fluid, the ensemble average of the entropic force is equilibrated by the average of the Stokes force that is proportional to the particle's velocity, 
\be\label{equi}
\left\langle F(r) \right\rangle = c\  \left\langle \frac{dr}{dt} \right\rangle ;
\ee 
for a spherical particle of radius $R$ the constant $c$ in the Stokes force is proportional to the shear viscosity of the fluid $\eta$:
\be\label{diff0}
c = 6 \pi R \eta .
\ee
In using the hydrodynamical notion of viscosity, we assume that the number of interactions needed to change $r$ substantially is very large. The ensemble average is thus performed over the continuous three-dimensional Gaussian probability distribution 
\be\label{prob_dist}
P(r) = \frac{4\ r^2}{\sqrt{\pi}\ q(t)^3} \ \exp\left(-\frac{r^2}{q(t)^2}\right) ,
\ee 
defined as follows: after time $t$ the particle will be located between $r$ and $r + dr$ with the probability $P(r) dr$ normalized by $\int P(r) dr = 1$, and $q(t)$ is the most probable value of $r(t)$. It is well known that the Gaussian distribution as a limit of Bernoullian distributions when the number of steps in a walk becomes very large \cite{Chandra}. 

The averages of different powers of $r$ over the distribution (\ref{prob_dist}) are given by 
\be\label{r2}
\langle r^2(t) \rangle = 3 q^2(t)/2 , 
\ee
\be\label{r1}
\langle r(t) \rangle = 2 q(t)/\sqrt{\pi} , 
\ee
\be\label{r-1}
\langle 1/r(t) \rangle = 2 /(\sqrt{\pi} q(t)) . 
\ee
 Substituting (\ref{r1}) and (\ref{r-1}) into 
(\ref{equi}) and using the expression for the entropic force (\ref{entf}), we get the differential equation
\be\label{diff}
q dq = 2 D dt ,
\ee
where $D$ is the diffusion coefficient that according to (\ref{diff0}) is given by 
\be
D = \frac{kT}{c} = \frac{kT}{6 \pi R \eta} .
\ee
The solution of (\ref{diff}) consistent with the initial condition $q(t=0) =0$ is 
\be
q^2(t) = 4 D t .
\ee
Using $\langle x^2 \rangle = \langle r^2 \rangle/3$ for isotropic diffusion in three spatial dimensions, and $q^2 = 2r^2 /3 = 2 x^2$, we get the Einstein relation for diffusion:
\be\label{einstein}
\langle x^2(t) \rangle = 2 D t .
\ee
\vskip0.2cm
\section{The Chandrasekhar's law} 

Let us now consider the particle bound to the origin by a quadratic potential $U(r) =  a r^2/2$ resulting in the Hooke's force 
\be\label{linforce}
F_H = - \frac{\partial U}{\partial r} = - a r .
\ee
Equating the average of the force $F_H$ to the average of the entropic force (\ref{entf}) (pointing in the opposite direction) similarly to (\ref{equi}), we get
\be
a \langle r \rangle = 2 k T \left\langle \frac{1}{r} \right\rangle .
\ee
Using (\ref{r1}), (\ref{r-1}), and $q^2 = 2 \langle x^2 \rangle$ we get
\be
\langle x^2 \rangle = \frac{k T}{a} .
\ee
This is the classic Chandrasekhar's law \cite{Chandra} underlying the theory of thermal expansion. 
Note that this derivation based on the entropic force \cite{Neumann} is significantly simpler than the original one \cite{Chandra}. 
\vskip0.2cm
If we consider the force (\ref{linforce}) as resulting from the interaction among the constituents of a bound state, then we can note that
unlike in the case of diffusion (\ref{einstein}), the distance between the constituents does not increase with time, so the state does not dissolve. However the square of the effective size of the bound state  grows linearly with temperature. 
\vskip0.2cm
\section{The law for linear confinement}
 
Let us now assume a linear confining potential $U(r) = \sigma r$ with a string tension $\sigma$; the corresponding force is  $F_c = - \partial U/\partial r = - \sigma$. The balance of the confining and entropic forces yields
\be
\sigma = 2 k T \left\langle \frac{1}{r} \right\rangle .
\ee
Using (\ref{r-1}) we get
\be
q^2(t) = \frac{16}{\pi} \frac{(k T)^2}{\sigma^2} ;
\ee
since $\langle x^2 \rangle = q^2/2$ we find for the average distance  
\be\label{linear}
\langle x^2 \rangle  = \frac{8}{\pi} \left(\frac{k T}{\sigma}\right)^2 .
\ee
We thus find that, in analogy with Chandrasekhar's law, the square of the average distance between the constituents grows with temperature, but for the linear confining potential the dependence on the temperature is {\it quadratic}. This means that hadronic systems bound by the confining potential undergo a much more pronounced "thermal expansion" than the ones governed by the Chandrasekhar's law.
\vskip0.2cm
Let us investigate the consequences of the relation (\ref{linear}) for the dissociation of quarkonium in quark-gluon plasma. As the temperature $T$ of the plasma grows, the average distance between the heavy quark and antiquark will increase, and at some value $T=T_d$ will reach the distance $x_s(T)$ at which the potential is screened and the entanglement entropy no longer depends on the distance, so the quarks become uncorrelated. It is natural to associate this temperature $T_d$ with the dissociation temperature at which the heavy quarkonium "melts". 
The string tension also has a mild dependence on the temperature $\sigma = \sigma(T)$.  
We thus get from (\ref{linear}) an equation for the dissociation temperature:
\be\label{lineq}
\langle x_s^2(T) \rangle  = \frac{8}{\pi} \left(\frac{k T}{\sigma(T)}\right)^2 .
\ee
The screening length $x_s(T)$ and the string tension $\sigma(T)$ have been extensively studied in Euclidean lattice QCD simulations, see e.g. \cite{Kaczmarek:2002mc,Petreczky:2004pz,Kaczmarek:2005zp}; using this input, we can solve (\ref{lineq}) and find the dissociation temperature $T_d$. Using the lattice data from \cite{Kaczmarek:2005zp}, we get an estimate of the dissociation temperature, $T_d \simeq 300$ MeV. However, as we will now discuss, this estimate misses a very important feature of the lattice data -- a substantial entanglement of the heavy quark pair with the quark-gluon plasma, and the entropy associated with it \cite{Kaczmarek:2002mc,Petreczky:2004pz,Kaczmarek:2005zp}.
\vskip0.2cm

\section{Entropic self-destruction} 

The lattice QCD data clearly indicate the presence of a significant additional entropy associated with a static heavy quark-antiquark pair \cite{Kaczmarek:2002mc,Petreczky:2004pz,Kaczmarek:2005zp}, see Fig.{\ref{lattice}.  Moreover, in a broad range of quark-antiquark distances $r$, 
this entropy $S=S(r, T)$ increases linearly in $r$, indicating an {\it exponential growth} of the number of states $\Omega(r)$ with the distance. 
The exponential growth of the number of states is in sharp contradiction with the conventional power law $\Omega(r) \sim r^2$ (see (\ref{entdif})) that drives the usual diffusion described by Einstein's law (\ref{einstein}). 
This behavior may result from the presence of delocalized "long string" states that possess large entropy far exceeding that of a two-particle quark-antiquark state, see Fig. \ref{condensate}. In other words, the quark-antiquark pair is strongly entangled with the rest of the system. 
\vskip0.2cm

\begin{figure}[h]
\includegraphics[width =9cm]{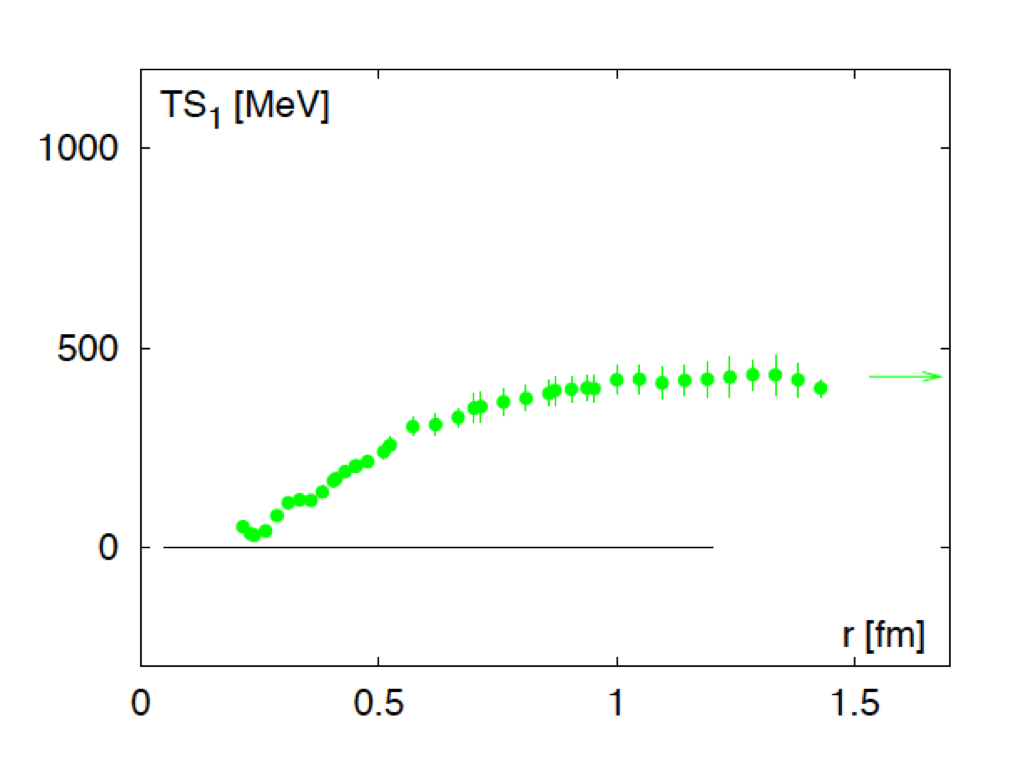}
\caption{\label{lattice} The additional entropy $S_1$ (multiplied by temperature $T$) associated with the color singlet quark-antiquark state at temperature $T \simeq 1.3\ T_c$ ($T_c$ is the deconfinement temperature) in 2 flavor lattice QCD as a function of the distance between the quark and antiquark; from \cite{Kaczmarek:2005zp}.  }
\end{figure}

Within the range of distances where the entropy is approximately linear in $r$ (see Fig.{\ref{lattice}) $S(r, T) = k s^\prime(T) r + const$, the entropic force 
\be\label{linen}
F = T \frac{\partial S(r, T)}{\partial r} = k T s^\prime(T)
\ee
does not fall off with the distance unlike (\ref{entf}) and is thus much more efficient in dissociating the bound states. 
\vskip0.2cm
At short distances $r$, the quark and antiquark represent a small color dipole and decouple from long wavelength gluon excitations. Because of this, the entanglement of the pair with the rest of the system is small, the corresponding entropy as indicated by Fig.{\ref{lattice} vanishes, and the heavy quarkonium is intact.  In this regime the dominant mechanism of heavy quarkonium dissociation is by the impact of thermal gluon fluctuations \cite{Kharzeev:1994pz} through the QCD version of photo-effect \cite{Shuryak:1978ij,Bhanot:1979vb}.
On the other hand, at large distances the quark and antiquark are no longer correlated and the entropy no longer depends on $r$. In this regime the entropic force is dominated by the conventional expression (\ref{entdif}), and the motion of heavy quarks is driven by the Einstein's diffusion (\ref{einstein}).  
\vskip0.2cm
The difference in the dissociation mechanisms in the hadron gas at $T < T_c$ and in the deconfined phase around $T_c$ is illustrated in Fig. \ref{condensate}. In the hadron gas phase, the confining interaction between the heavy quark and antiquark is screened by the produced light quark-antiquark pair, leading to the production of two open charm mesons (see left panel of Fig. \ref{condensate}). In this case the number of physical states can be expected to grow as a square of the distance between the heavy quark and antiquark, similar to the case of diffusion (\ref{entdif}). 
In the deconfined phase, the number of physical states grows exponentially with the inter-quark distance $r$, corresponding to the linear increase of entropy with $r$ observed at intermediate values of $r$, see Fig.{\ref{lattice}. This exponential growth likely originates from coupling to the "long string" excitations that are characterized by a large density of states, see the right panel of Fig. \ref{condensate}. 
\vspace{-0.3cm}
\begin{figure}[h]
\includegraphics[width =8.5cm]{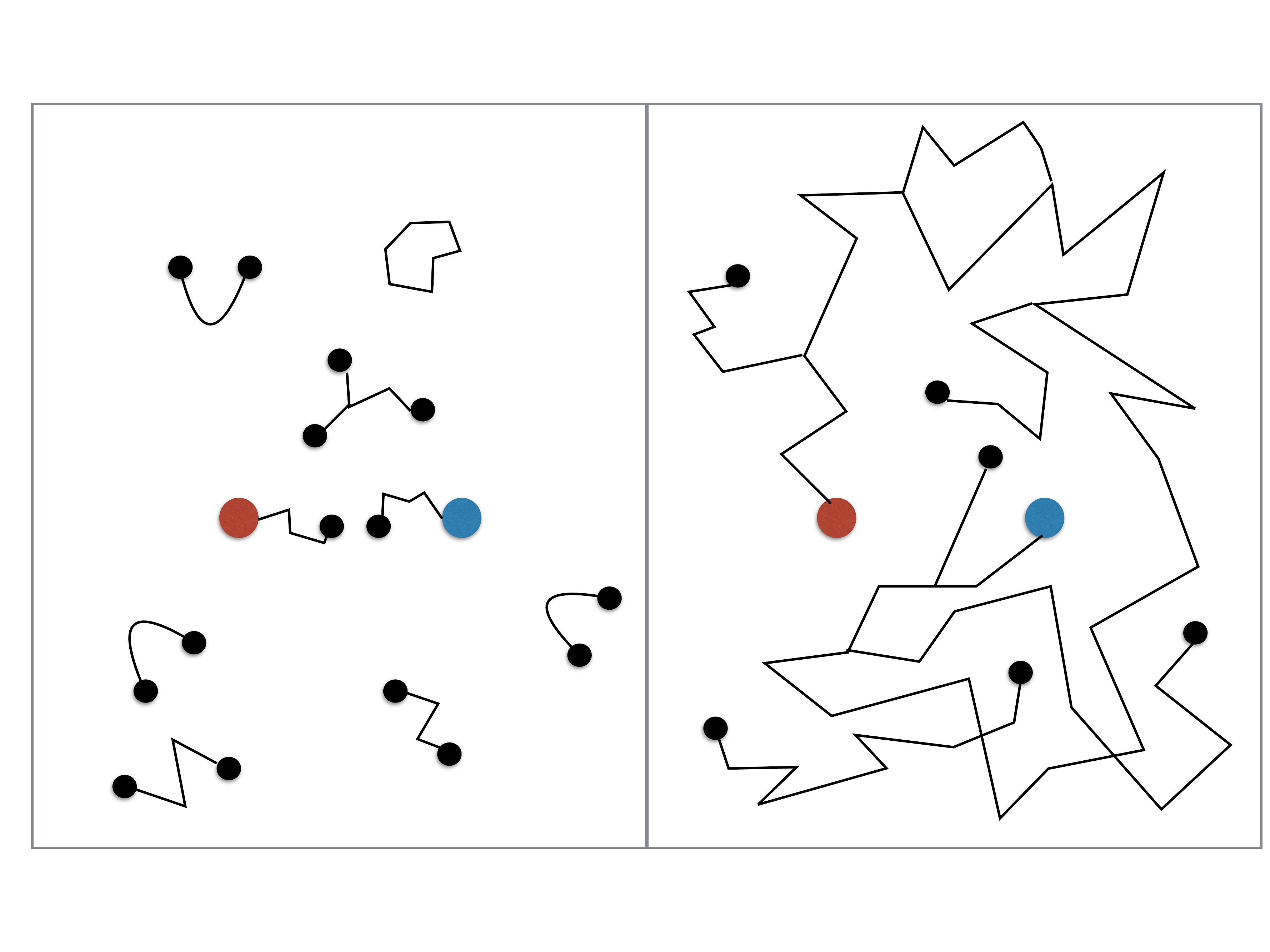}
\caption{\label{condensate} The dissociation of heavy quark-antiquark state in the hadron gas (left panel) and in the deconfined phase near the transition temperature (right panel).}
\end{figure}
\vskip0.2cm
The presence of the entropy $S$ associated with the quark pair means that the free energy $A$ and internal energy $U$ differ, since $A = U - T S$. It has been proposed to use the free energy \cite{Digal:2001ue}, the internal energy \cite{Shuryak:2004tx,Alberico:2005xw}, or combination of the two \cite{Wong:2005be} as inputs in potential model calculations of heavy quarkonium spectra in the medium. In the present author's opinion, the entropy associated with the quark pair signals that the pair couples to on-shell degrees of freedom, and this coupling invalidates the key assumption of the potential approach - namely, that the interaction between the constituents is instantaneous. The coupling to on-shell degrees of freedom (manifested by the entropy) inevitably introduces retardation effects and leads to the breakdown of the potential model. We thus need a different treatment taking account of the entropy.

\vskip0.2cm
In the lattice setup, the quarks are static, and the measured entropy (let us call it $S_{lat}(r,T)$) does not include the entropy (\ref{entf}) resulting from the quark motion in coordinate space. Since the entropy is additive, the total entropic force is thus given by the sum
\be\label{toten}
F(r) = T \frac{\partial S_{lat}(r, T)}{\partial r} + \frac{2 k T}{r} .
\ee
The balance of the average of (\ref{toten}) and the confining force yields
\be\label{balen}
T \frac{\partial S_{lat}(r, T)}{\partial r} + \frac{4 k T}{\sqrt{\pi} q} = \frac{\partial U (r, T)}{\partial r} ,
\ee
where we used (\ref{r-1}). 
\vskip0.2cm
Let us assume that both the entropy and the confining potential are linear in $r$, $S_{lat}(r, T) = k s^\prime(T)\ r + const$ and $U(r, T) = \sigma(T)\ r$, as indicated by the lattice data at intermediate distances $r$. Using $q^2 = 2 x^2$ we get from (\ref{balen})
\be\label{fin}
\langle x^2 \rangle  = \frac{8}{\pi} \left(\frac{\sigma(T)}{k T} - s^\prime(T)\right)^{-2} .
\ee 
If we neglect the entropy $S_{lat}$ describing the entanglement of quarkonium with the plasma and put $s^\prime(T)=0$ in (\ref{fin}), we recover the law (\ref{linear}). The dependence of $\langle x^2 \rangle$ on the temperature as given by (\ref{fin}) is illustrated on Fig. \ref{Fig.2}; to produce this plot, we assumed for simplicity the fixed values of  $\sigma \simeq 0.2$ GeV$^2$ and $s^\prime \simeq 0.7$ GeV.
\vskip0.2cm
\begin{figure}[h]
\includegraphics[width =9cm]{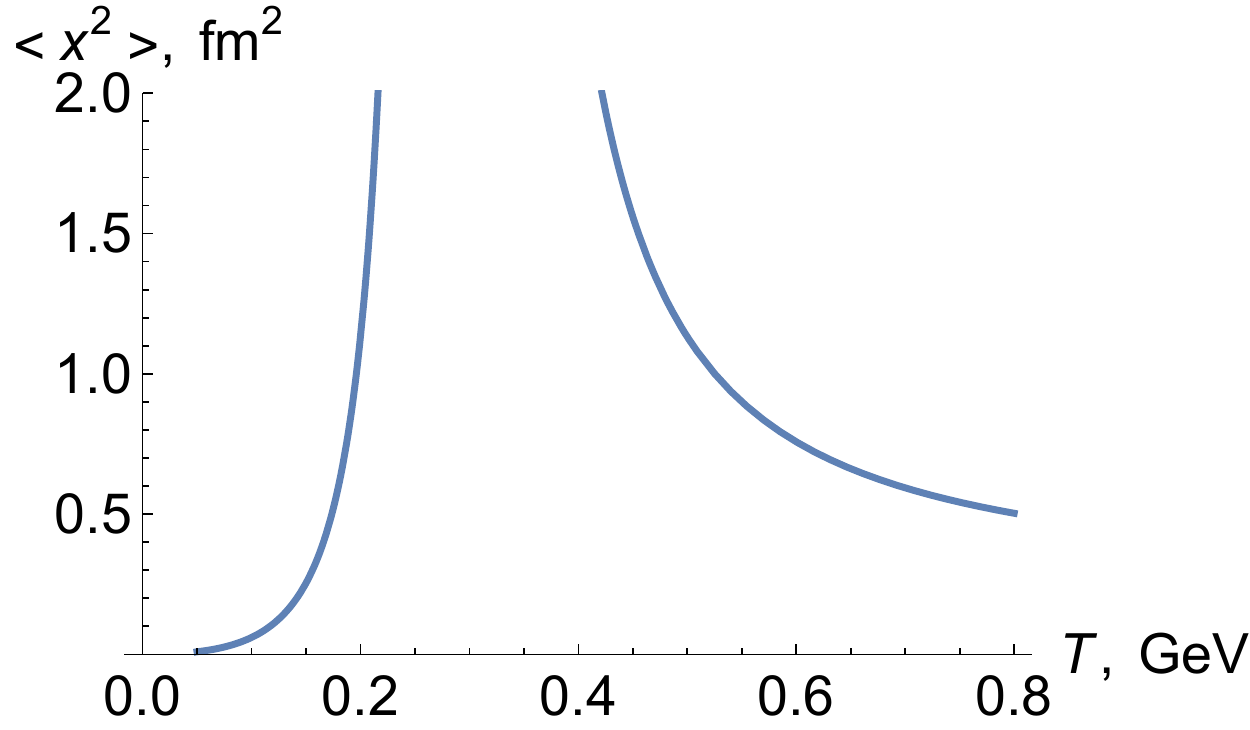}
\caption{\label{Fig.2} The mean radius squared $\langle x^2 \rangle$ of the heavy quark-antiquark bound state in strongly coupled quark-gluon plasma as a function of temperature $T$, as given by (\ref{fin}). The value of $(\langle x^2 \rangle)^{1/2}$ that exceeds the screening length implies the dissociation of the bound state.}
\end{figure}

It is clear from (\ref{fin}) that the entanglement entropy leads to a dramatic increase of the average distance between the heavy quarks. In particular, when 
\be\label{deloc}
s^\prime(T) = \frac{\sigma(T)}{k T},
\ee
the quark-antiquark state becomes completely delocalized! 
\vskip0.2cm


The condition for dissociation, as before, is $\langle x^2 \rangle \geq \langle x_s^2(T) \rangle$, where $x_s(T)$ is the screening distance at which the confining potential is screened and the entanglement entropy no longer depends on the distance. When the condition (\ref{deloc}) is met, $\langle x^2 \rangle \to \infty$, and no bound states 
exist {\it in equilibrium}. 
\vskip0.2cm
In two flavor QCD as in Fig. \ref{lattice}, $k T_c \simeq 200$ MeV \cite{Kaczmarek:2005ui}, so the divergence of the relative distance occurs at 
\be
k T = \frac{\sigma(T)}{s^\prime(T)} \simeq 280\ {\rm MeV} \simeq 1.4\ k T_c .
\ee
However the average distance starts to exceed the screening length already around $T_c$, as can be seen from Fig. \ref{Fig.2}.
This means that around $T_c$ {\it all} bound hadronic states should cease to exist.   It is thus tempting to speculate that the condition (\ref{deloc}) presents an entropic view on the deconfinement itself. Namely, the deconfinement occurs because the excited hadron states become delocalized and entangled. This resembles the percolation picture of deconfinement \cite{Celik:1980td,Satz:1998kg}, in which the size of the percolation cluster diverges at the deconfinement phase transition. 
\vskip0.2cm
Our assumption of the linear dependence of the entropy on the inter-quark distance $r$ holds only within the range $0.2\ {\rm fm} < r < 0.6\ {\rm fm}$, see 
Fig. \ref{lattice}; at larger distances, the entropy flattens off. However it is this range of distances which is crucial for our considerations, since once the inter-quark distance exceeds the screening length the quarkonium dissociates. At inter-quark distances exceeding the screening length both the entropy and internal energy cease to depend on $r$, and we get back to the Einsten diffusion of heavy quarks in the plasma described by (\ref{einstein}).   
\vskip0.2cm
  
It is nevertheless interesting to examine the cases when the entropy's dependence on the inter-quark distance is slower than linear, e.g. a) logarithmic $S(r) = a \ln r$ or b) square root $S(r) = a' \sqrt{r}$. Repeating the computations made above, we find that in the case a) the square of the inter-quark distance $<x^2>$ grows quadratically with the ratio of temperature to string tension, similarly to (\ref{linear}) but with a larger coefficient $(2a+4)^2/2\pi$; 
when $a \to 0$, we recover $8/\pi$ as in (\ref{linear}). 
For the case b), we need the average of $1/\sqrt{r}$ over the distribution (\ref{prob_dist}); it is given by
\be
\left\langle \frac{1}{\sqrt{r(t)}} \right\rangle = \frac{2 \Gamma\left(\frac{5}{4}\right)}{\sqrt{\pi q(t)}} . 
\ee
The resulting expression for $<x^2>$ is easily obtained by solving a quadratic equation for $\sqrt{q}$; it has simple low and high temperature $T$ limits. At low $T$, we recover (\ref{linear}).  In high $T$ limit, we get
\be
\left< x^2 \right> = \frac{a'^4 \Gamma\left(\frac{5}{4}\right)^4}{2 \pi^2} \left(\frac{k T}{\sigma}\right)^4 ,
\ee
which is quartic in temperature and thus signals a much faster increase of the thermal expansion than (\ref{linear}).
\vskip0.2cm
We are now ready to address the heavy quarkonium suppression puzzle. 
The key lattice observation in this case is the following: the additional entropy associated with the heavy quark pair peaks around $T_c$, and essentially vanishes above $1.5\ T_c$ \cite{Kaczmarek:2005zp}. Since the entropic force drives the dissociation process in our scenario, the suppression of charmonia (which have the sizes affected by the presence of the entropy) has to be stronger at temperatures close to $T_c$ (which is the case at RHIC energy) than at higher temperatures achieved at the LHC. On the other hand, most of the bottomonium states have smaller sizes, and are thus much less affected by the entropic forces. In accord with our discussion above, this implies that their dissociation mechanism is mostly conventional (Debye screening \cite{Matsui:1986dk} or thermal gluon activation \cite{Kharzeev:1994pz}) leading to a sequential suppression pattern \cite{Karsch:2005nk}, and thus the bottomonium suppression gets stronger at higher energy densities. The available data indicate that the bottomonium suppression is indeed stronger at the LHC \cite{Chatrchyan:2012lxa,Abelev:2014nua} than at RHIC \cite{Adamczyk:2013poh,Adare:2014hje}, in accord with our scenario.
\vskip0.2cm

Of course, a detailed quantitative study including state-of-the art analysis of the available lattice QCD results and a real time evolution of the quark-gluon plasma is needed to reach a definitive conclusion. This study is forthcoming, and will be presented elsewhere. Nevertheless, the entropic enhancement of charmonium dissociation at temperatures close to $T_c$  is a very robust feature of our scenario. Let us note also a similarity to the peak in jet quenching close to $T_c$ pointed out theoretically in \cite{Liao:2008dk} and indicated by the data on jet azimuthal distributions. 
 

\section{Conclusions}

The entanglement of a bound state with the rest of the system can lead to its entropic self-destruction. This happens in particular when the excitations of the bound state are delocalized and characterized by a large entropy. If this entropy increases with the separation between the constituents of the bound state, then the resulting entropic force} can drive the dissociation process. 
\vskip0.2cm
We have applied this treatment to the dissociation of heavy quarkonium in quark-gluon plasma, where lattice QCD indicates the presence of a large amount of entropy associated with the heavy quark pair, and 
this entropy grows with the inter-quark distance. We have argued that "entropic self-destruction" can lead to a strong suppression of the bound states close to $T_c$, possibly providing a solution to the heavy quarkonium suppression puzzle. A detailed quantitative study of this phenomenon will allow to check whether the proposed scenario adequately describes the experimental observations. 
\vskip0.2cm
The proposed mechanism of quarkonium dissociation underlines the importance of entanglement and entropy in the deconfinement transition. The presented approach to dissociation of bound states may also find applications in other systems with delocalized excitations, including topological phases in condensed matter. 
\vskip0.2cm

I am grateful to O. Kaczmarek, F. Karsch, S. Mukherjee, H. Satz, E. Shuryak, F. Wilczek, N. Xu, H.-U.Yee and W. Zajc for useful discussions. 
I thank the Alexander von Humboldt Foundation for supporting this work through the Research Award, and D. Rischke for warm hospitality at Goethe Universit${\rm\ddot{a}}$t in Frankfurt am Main and discussions.
  This work was supported in part by the U.S. Department of Energy under Contracts No.
DE-FG-88ER40388 and DE-AC02-98CH10886.


\end{document}